\documentstyle[12pt,epsf]{article}

\renewcommand{\bar}[1]{\overline{#1}}
\newcommand{\longvec}[1]{\,\mathrel{\rlap{\raise7pt
      \hbox{$\longrightarrow$}}\hbox{$#1$}}\,} 
\def\VEV#1{\left\langle{#1}\right\rangle}

\def\.{\hskip -4pt plus 10000pt}
\def\vr{\vrule width .6pt height 8pt depth 5pt} 
\def\srarrow{\rlap{\raise 8pt \hbox{\vr}}\. \rightarrow} 
\def\pslash{\not{\hbox{\kern-2.3pt $p$}}}
\def\sslash{\not{\hbox{\kern-4pt $s$}}}
\def\Wslash{\not{\hbox{\kern-4pt $W$}}}
\def\Sslash{\not{\hbox{\kern-2pt $S$}}}

\begin{document}
\begin{flushright}
SLAC--PUB--7360\\
November 1996
\end{flushright}
\bigskip\bigskip

\thispagestyle{empty}
\flushbottom
\vfill
\centerline{\large\bf Search For New Mechanism of CP Violation through}
\vspace*{0.035truein}
\centerline{{\large\bf Tau Decay and Semilpetonic
Decay of Hadrons}
    \footnote{\baselineskip=14pt
     Work supported by the Department of Energy, contract 
     DE--AC03--76SF00515.}}
\vspace{22pt}

  \centerline{\bf Yung Su Tsai}
\vspace{8pt}
  \centerline{\it Stanford Linear Accelerator Center}
  \centerline{\it Stanford University, Stanford, California 94309}
  \centerline{e-mail:ystth@slac.stanford.edu}
\vspace*{0.9cm}

\begin{center}
Abstract
\end{center}
If CP is violated in any decay process involving leptons it will
signify the existense of a new force (called the $X$ boson) responsible
for CP violation that may be the key to understanding matter-antimatter
asymmetry in the universe. We discuss the signatures of CP violation in
(1) the decay of tau lepton, and (2) the semileptonic decay of $\pi$,
$K$, $D$, $B$ and $t$ particles by measuring the polarization of the
charged lepton in the decay. We discuss how the coupling constants and
their phases of the coupling of the $X$ boson to 9 quark vertices and 3
lepton vertices can be obtained through 12 decay processes.

\vfill

\centerline{Presented to the}
\centerline{Proceedings of the Fourth International Workshop on Tau
Physics}
\centerline{Estes Park, Colorado}
\centerline{16--19 September 1996}
\vfill
\newpage

\section{INTRODUCTION}

The only particle that exhibits CP violation so far is the neutral $K$
\cite{Christenson} and in a few years we would know whether $B$ and $D$
mesons will have CP violation from the $B$ factories. The standard
theory of CP violation due to Kobayashi and Maskawa predicts
\cite{Kobayashi,Jarlskog} that there is no CP violation whenever lepton
is involved in the decay either as a parent or as a daughter. This
prediction applies for example to the decay of a muon and a tau lepton
or semileptonic decay of a hadron such as $\pi$, $K$, $D$, $B$ or $t$
as shown in Table \ref{tab1}. Since leptons constitute a sizable
fraction of the total number of particles in the universe this is a
statement of utmost importance and thus must be tested experimentally.
In order to have CP violation involving leptons we must go beyond the
standard model of Maskawa and Kobayashi. The reasons why in the
standard model leptons can not participate in CP violation are as
follows:

\begin{enumerate}
\item
Unlike $K_L$ decay into two pions, all the decays listed in Table
\ref{tab1} involve only a tree diagram of one $W$ exchange where $W$ is
coupled to each quark and leptonic vertex with one coupling constant in
the standard model.
\item
$T$ violation or CP violation occurs in quantum mechanics via existence
of a complex coupling constant in the vertex. The phase of this complex
coupling constant can not manifest itself if we have only $W$ exchange
diagram, we need another diagram to interfere with it to obtain its
phase. Hence $T$ or CP violation in the standard model can not occur in
the lowest order weak interaction.
\end{enumerate}
\noindent
Thus if we see CP  or $T$ violation in any of the decay processes shown
in Table 1 we can infer immediately the existence of another charged
boson mediating the weak interaction.  We shall call this new boson the
$X$ boson and the diagram involving the $X$ boson exchange $A_X$ and
the standard model diagram $A_W$ as shown in the figures. In Appendix B
we show that the $X$ particle must have spin zero. Since only the
relative phase of the two diagrams matters we shall assume that the
coupling constants appearing in $A_W$ are real and those appearing in
$A_X$ are allowed to be complex.

The only theoretical candidate for the $X$ particle is the charged
Higgs bosons proposed by Lee and Weinberg \cite{Lee,Weinberg}. The most
striking feature of its prediction is that the heavier the particle the
larger the CP violation. This should be tested experimentally. We shall
thus not assume that charged Higgs is the $X$ particle. This is to
avoid having a prejudice against testing CP violation involving lighter
particles. Both the Kobayashi-Maskawa model and the Lee-Weinberg Model
are based on the assumption that CP violation occurs spontaneously, and
this may not be true and must be tested experimentally. Tests of the
Standard Model involves testing the unitarity of the CKM matrix. If any
of the 6 unitarity triangles is found to be not closed then CKM model
of CP violation is wrong. If the sum of the absolute squares of three
elements in any row or column is not unity it will also show that CKM
matrix is not unitary. However to prove that all 12 conditions of
unitarity of the CKM matrix experimentally is an impossible task.

The vertex function for hadrons for the $X$ exchange diagram is related
to that for $W$ boson exchange diagram in the quark model, hence no new
form factor for $X$ coupling to hadron is needed. The purpose of this
paper is as follows:
\begin{enumerate}
\item
We give 12 examples of test of CP violation involving leptons in one
place so that experimentalists can choose a most suitable one to carry
out the experiment. After one experiment is found to have $T$ or CP
violation, there will be an onrush of effort to finish all the
experiments in the table. We will then know the coupling constants
between $X$ boson and all the quarks and leptons. We will also know
approximately the mass of the $X$ boson.
\item 
We have avoided purposely to assume that the $X$ boson is the charged
Higgs boson of Lee and Weinberg \cite{Lee,Weinberg} so that
experimenters will not have prejudice to avoid testing CP violation for
light particles. By avoiding a discussion of the origin of complex
phases in the coupling constants for X particles, we can concentrate on
model independent features of CP violation such as the role of CPT
theorem, the role of complex phases due to final state interactions,
the role played by the complex part of the $W$ propagator when it is on
the mass shell such as in the top decay, and the use of polarization in
the initial state and the final states to obtain CP violating effects.
We also discuss test of CP using partially integrated cross section
without using polarizations.

\item
The $X$ boson could be the particle we need for causing the
matter-antimatter asymmetry in our universe \cite{Turok,Dine}.
\end{enumerate}

\begin{table*}
\caption[*]
{Test of CP and T Violations Involving Leptons. (Test of Existence of a
New Boson Responsible
for CP Violation.) Determination of Coupling of $X$ Boson to 12 Vertices.}
\label{tab1}
{\scriptsize
\begin{center}
\begin{tabular}{|cllll|}\hline\hline
Vertex
& Experiment(where)
& Signature of CP, T or CPT violation
& Obtains
& Ref.
\\ \hline
$X_{ud}$
& 1.\ $\pi^+ \rightarrow \pi^0+ e^+ +\nu\ (LA)$ & $C_1 (p_{\pi^+}\times
p_{\pi^0})\cdot
W_{e^+}\ \ \ \ \ \ C_1 \ne 0\ (T)$
& ${\rm Im}(X_{ud} X_{e \nu})$
& [9] \\
$X_{us}$
& 2.\ $\tau^\pm \rightarrow K^\pm+\pi^0+\nu\ (TCF)$ 
& $C_2^\pm(p_{K^\pm}\cdot
W_{\tau^\pm}) \ \ \ \ \ \ C_2^+\ne-C_2^- \ (CP)$ &${\rm Im}(X_{us}
X_{\tau\nu})$ & [10]
\\
& &
$C_3^\pm(p_{\pi^0}\cdot W_{\tau^\pm})$
\ \ \ \ \ \ \ $ C_3^+\ne -C_3^-\ (CP)$
& ${\rm Im}(X_{us} X_{\tau\nu})$
&
\\
& &
$C_4^\pm(p_{\pi^0}\times p_{K^\pm})\cdot W_{\tau^\pm} 
$ \ \ \ \ $C_4^+\ne
C_4^- (CP)$
& ${\rm Im}(X_{us} X_{\tau\nu})$&
\\
& 3.\ $K^\pm\rightarrow\pi^0+\mu^\pm+\nu\ (BNL)$ 
& $C_5^\pm(p_{\pi^0}\times
p_{\mu^\pm})\cdot W_{\mu^\pm}\ \ \ \ C_5^\pm \ne 0 \ (T)$ 
& ${\rm Im}(X_{us}
X_{\mu\nu})$ & [11-13]
\\
$X_{ub}$
& 4.\ \ $B^\pm\rightarrow\pi^0+\tau^\pm+\nu\ (BF)$ 
& $C_5^\pm(p_{\pi^0}\times
p_{\tau^\pm})\cdot W_{\tau^\pm}\ \ C_5^+ \ne -C_5^- (CPT) $ 
& ${\rm Im}(X_{ub}
X_{\tau\nu}) $ & [14]
\\
$X_{cd}$
& 5.\ \ $D^\pm\rightarrow\pi^0+\mu^\pm+\nu\ (TCF)$ 
& $C_5^\pm(p_{\pi^0}\times
p_{\mu^\pm})\cdot
W_{\mu^\pm}$ & ${\rm Im}(X_{cd} X_{\mu\nu})$ & [14]
\\
$X_{cs}$
& 6.\ \ $D^\pm\rightarrow K^0+\mu^\pm+\nu\ (TCF)$ 
& $C_5^\pm(p_{K^0}\times
p_{\mu^\pm})\cdot W_{\mu^\pm}$ & ${\rm Im}(X_{cs} X_{\mu\nu})$ & [14]
\\
$X_{cb} $
& 7.\ \ $B^\pm\rightarrow D^0+\tau^\pm+\nu\ (BF)$ 
& $C_5^\pm(p_{D^0}\times
p_{\tau^\pm})\cdot W_{\tau^\pm}$ & ${\rm Im}(X_{cb} X_{\tau\nu})$ & [14]
\\
$X_{td} $

& 8.\ \ $t^\pm\rightarrow \pi^0+\tau^\pm+\nu\ (FNL)$ 
& $C_5^\pm(p_{\pi^0}\times
p_{\tau^\pm})\cdot W_{\tau^\pm}$ & ${\rm Im}(X_{td} X_{\tau\nu})$ & [15]
\\
$X_{ts} $
& 9.\ \ $t^\pm\rightarrow K^0+\tau^\pm+\nu\ (FNL)$ 
& $C_5^\pm(p_{K^0}\times
p_{\tau^\pm})\cdot W_{\tau^\pm}$ & ${\rm Im}(X_{ts} X_{\tau\nu})$ & [15]
\\
$X_{tb} $
& 10.\ \ $t^\pm\rightarrow B^0+\tau^\pm+\nu\ (FNL)$ 
& $C_5^\pm(p_{B^0}\times
p_{\tau^\pm})\cdot W_{\tau^\pm}$ & ${\rm Im}(X_{tb} X_{\tau\nu})$ & [15]
\\
$X_{e\nu}$
& 11.\ \ $\mu^\pm\rightarrow e^\pm+\nu_\mu+\nu_e\ (LA)$ &
$C_6^\pm(W_{e^\pm}\times
p_{e^\pm})\cdot W_{\mu^\pm}\ \ \ \ C_6^\pm \ne 0 \ (T)$ 
& ${\rm Im}(X_{\mu\nu}
X_{e\nu})$ & [16]
\\
$X_{\mu\nu}$
& 12.\ \ $\tau^\pm\rightarrow \mu^\pm+\nu_\mu+\nu_\tau\ (TCF)$ &
$C_6^\pm(W_{\mu^\pm}\times
p_{\mu^\pm})\cdot W_{\tau^\pm}\ \ C_6^+ \ne C_6^- (CPT)$ &
${\rm Im}(X_{\mu\nu} X_{\tau\nu})$ & [16] \\ $X_{\tau\nu}$
& \multicolumn{2}{l}{Obtainable from Experiment\ 
1, 2, 4, 7, 8, 9, 10 or
12.} &&\\
\hline \hline \end{tabular}
\end{center}}
\end{table*}

In Section 2 we treat the semileptonic decay of Tau using $\tau^\pm
\rightarrow \nu_\tau + K^\pm+\pi^0$ as an example. This is the only
case in Table \ref{tab1} that involves final state interaction. We show
that only the interference term between $s$ wave from $X$ exchange and
$p$ wave from $W$ exchange diagrams can produce CP violation. In
Section 3 we treat a simple semileptonic decay of a hadron. We chose
spin zero hadrons for initial and final state because they are simplest
to analyze. We also chose final hadrons to be neutral so that we do not
have to worry about corrections due to electromagnetic final state
interactions. Since there is no final state interaction, we can have
only the $T$ odd term to look for CP violation. The only  $T$ odd term
in the problem is $(\bf {p_3 \times p_4}) \cdot {\bf W}$, where $\bf
p_3$, $\bf p_4$ and $\bf W$ are the incident hadron momentum, the
outgoing hadron momentum and the lepton polarization, all measured in
the rest frame of the lepton.

Nelson et al. \cite{Nelson} were the first ones to consider the
possibility of CP violation in the semileptonic decay of tau lepton.
However their treatments of the problem violates many sacred principles
such as TCP invariance and rotational invariance (see Appendix A).

\section{SEMILEPTONIC DECAY OF TAU}

\begin{figure}[htb]
\leavevmode
\centering
{\epsfxsize=2.9in\epsfbox{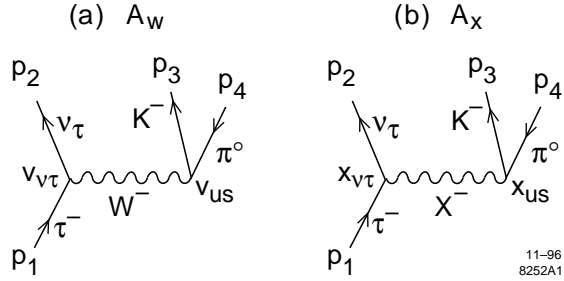}}
\caption[*]{Feynman diagrams for the reaction $\tau^-\rightarrow
\nu_\tau + K^-+\pi^0$. $A_W$ is the Standard Model $W^\pm$ exchange
diagram. $A_X$ is the CP violating scalar exchange, $X^\pm$ may or may
not be the charged Higgs boson.}
\label{fig1}
\end{figure}

Let us consider the decay $\tau^-\rightarrow K^-+\pi^0+\nu$
(see Fig. 1). The CP
conserving $W$ exchange
diagram of the Standard Model can be written as 
\begin{equation}
A_W = V_{us}[(p_3-p_4)_\mu f_-(s)+(p_3+p_4)_\mu f_+(s)] 
\bar u(p_2)\gamma_\mu(1-\gamma_5)u(p_1) \label{eq:2.1}
\end{equation}
where $V_{us}$ is the CKM matrix which can be chosen to be real. $p_3$,
$p_4$, $p_2$, and $p_1$ are four momenta of $K^-$, $\pi^0$, $\nu_\tau$
and $\tau$ respectively, and $s = (p_3+p_4)^2$.

The hadronic current in Eq. (\ref{eq:2.1}) is a pure vector (not an
axial vector) because $K^-\pi^0$ can only be $J^P=0^+$ and $1^-$. Hence
\begin{equation}
J_\mu \equiv
\VEV{K^-\pi^-\,|\,\bar\psi_u(0)\gamma_\mu\psi_s(0)\,|\,0} =
(p_3-p_4)_\mu
f_-(s)+(p_3+p_4)_\mu f_+(s) \label{eq:2.2}
\end{equation}

In {\it the rest frame} of $p_3+p_4$ the zeroth component of $J_\mu$,
denoted by $J_{ro}$ which is rotationally invariant, represents the $s$
wave and the vector part $\vec J_r$ represents the $p$ wave. The $s$
wave part is thus
\begin{equation}
J_{ro} =
\left[\frac{m^2_3-m^2_4}{\sqrt s}\, f_-(s)+\sqrt s\, f_+(s) \right]
\label{eq:2.3}
\end{equation}
and the $p$ wave part is
\begin{equation}
\vec J_r = \left[2\vec p_{3r}\, f_-(s)\right] \ , \label{eq:2.4}
\end{equation}
where $\vec p_{3r} = - \vec p_{4r}$ is the momentum of $K^-$ in the
rest frame of the $K^-+\pi^0$ system. The amplitude for the CP
nonconserving $X$ scalar boson exchange is
\begin{equation} 
A_X = CX_{us}X_{\tau\nu}\, J_{ro}\, \bar u(p_2)(1+\gamma_5)\, u(p_1)
\label{eq:2.5}
\end{equation}
where $X_{us}$ and $X_{\tau\nu}$ are the complex coupling constants. 
The proportionality constant $C$ can be obtained from
\begin{equation} 
\VEV{K^-\pi^0\,|\,\bar\psi_d(0)\psi_s(0)\,|\,0} =
CJ_{ro} = \frac{(p_3+p_4)_\mu}{m_s-m_u}\, J_\mu = 
\frac{\sqrt s}{m_s-m_u}\, J_{ro}
\label{eq:2.6}
\end{equation}
and hence
\begin{equation}
C = \frac{\sqrt s}{m_s-m_u} \ .
\label{eq:2.7}
\end{equation}
From Eq. \ref{eq:2.5}, \ref{eq:2.6} and \ref{eq:2.7}, we have 
\begin{equation}
A_X = X_{us}X_{\tau\nu}\, \frac{s}{m_s-m_u}\, f_0\bar u(p_2)
(1+\gamma_5)u(p_1) \ ,
\label{eq:2.8}
\end{equation}
where
\begin{equation}
f_0(s) = \frac{{M_3}^2-{M_4}^2}{s}\, f_-(s)+f_+(s) \label{eq:2.9}
\end{equation}
is the $s$ wave form factor and its phase is defined by 
\begin{equation}
f_0(s) = |f_0(s)|e^{i\delta_0(s)}\ .
\label{eq:2.10}
\end{equation}
Only the $p$ wave part of $A_W$ interfering with $A_X$ can produce CP
violation \cite{Tsai95a,Tsai96a,Tsai96b}. Therefore instead of $f_-(s)$
and $f_+(s)$ in Eq. (\ref{eq:2.1}) we write
\begin{eqnarray}
A_W &=& V_{us} \Big[ f_1(s)
\left\{(p_3-p_4)_\mu-\frac{{M_3}^2-{M_4}^2}{s}\,(p_3+p_4)_\mu\right\} 
\nonumber \\ &&
+ f_0(s)(p_3+p_4)_\mu\Big]\, \bar u(p_2) \gamma_\mu(1-\gamma_5)u(p_2)\,
\label{eq:2.11}
\end{eqnarray}
where $f_1(s) = f_-(s)$ is the $p$ wave form factor, 
and its phase if defined by
\begin{equation}
f_1(s) = |f_1(s)|e^{i\delta_1(s)} \ .
\label{eq:2.12}
\end{equation}
The decay probability of $\tau^-\rightarrow \nu_\tau+K^-+\pi^0$ can be
written as
\begin{equation}
\Gamma = \frac{1}{2M_\tau}\, \frac{1}{(2\pi)^5}
\int \frac{d^3p_2}{2E_2} \int
\frac{d^3p_3}{2E_3}\int \frac{d^3p_4}{2E_4} 
\delta^4(p_1-p_2-p_3-p_4)
\left|\frac{g^2}{{m_W}^2}\, A_W+ \frac{1}{{m_X}^2}
\, A_X\right|^2 
\label{eq:2.13}
\end{equation}
where $g^2/m^2_W = G/\sqrt 2$ with $G = 1.116\times 10^{-5}\ GeV^{-2}$,
hence $g = 0.2303$. Let
\begin{eqnarray*}
V &=& f_0(\pslash_2+\pslash_4) + f_1
\left\{\pslash_3-\pslash_4-D(\pslash_3+\pslash_4)\right\}, \\[2ex]
V^*&=& f^*_0 (\pslash_3+\pslash_4)+
f^*_1\left\{\pslash_3-\pslash_4- D(\pslash_3+\pslash_4)\right\}.
\label{eq:2.14}
\end{eqnarray*}
We have
\begin{eqnarray}
\frac{A^+_WA_W}{2|V_{us}|^2} &=& \frac{{\rm Tr}}{4}\, 
\Big[(1+\gamma_5 \Wslash)(\pslash_1+M_1)
V(1-\gamma_5)\pslash_2(1+\gamma_5) V^*\Big] \nonumber \\
&=& |f_0|^2 \Big\{ W_3(-2M^3_1)+W_4(-2M^3_1)
+M^2_1(-s+M^2_1)\Big\}
\nonumber \\ && + |f_1|^2 \Big\{ W_32M_1
                \Big[2Q\cdot p_1(D-1)-D^2M^2_1
+D(s+M^2_1)-s-2(M^2_3+M^2_4)\Big]
\nonumber \\ && + W_42M_1\Big[ 2Q\cdot p_1(D+1)-D^2M^2_1
-D(s+M^2_1)+s-2(M^2_3+M^2_4)\Big] 
\nonumber \\ && + \left[ 2Q\cdot p_1-D(s+M^2_1)\right]^2 
+ (s-M^2_1)(2M^2_3+2M^2_4-s-D^2s)\Big\} 
\nonumber \\ &&+ 2{\rm Re}(f_1f^*_1)\Big\{ W_3M_1(-2Q\cdot p_1 
+2DM^2_1+s-M^2_1) 
\nonumber \\ && +W_4 M_1 (-2Q\cdot p_1+2DM^2_1-s+M^2_1)
+M^2_1[2Q\cdot p_1-D(s+M^2)1]\Big\} 
\nonumber \\ && - 8M_1 {\rm Im} (f_0f^*_1)Eps (W, p_3, p_4,p_1) \ , 
\label{eq:2.15}
\end{eqnarray}
where $W_3 = W\cdot p_3,\ W_4=W\cdot p_4,\ D=(M^2_3-M^2_4)/s,\ s =
(p_3+p_4)^2, Q = p_3-p_4$;
\begin{eqnarray}
&&\frac{A^+_WA_X+A^+_XA_W}{2|V_{us}X_{us}X_{\tau\nu}|} 
= \frac{{\rm
Tr}}{4}\, (1+\gamma_5 \Wslash)(\pslash_1+M_1) 
\Big[(1-\gamma_5)\pslash_2(1+\gamma_5)
\nonumber \\ && V^*f_0
+V(1-\gamma_5) \pslash_2(1+\gamma_5)f^*_0\Big]
\nonumber \\ &=& \cos(\delta_t)|f_0|
\left\{-4W_3M^2_1-4W_4M^2_1+2M_1(-s+M^2_1)\right\}
+ 2\cos(\delta_0-\delta_1+\delta_t)|f_1|
\nonumber \\ && \Big\{W_3(-2Q\cdot p_1+2DM^2_1-s-M^2_1) 
+ W_4(-2Q\cdot p_1+2DM^2_1-s+M^2_1)
\nonumber \\ && +M_1[2Q\cdot p_1-D(s+M^2_1]\Big\}
-8|f_1|\sin (\delta_0-\delta_1+\delta_t) 
                Eps (W, p_3,p_4,p_1) \ ,
\label{eq:2.16}
\end{eqnarray}
where $\delta_t $ is the CP violating phase in $A_X$ , i.e.
$X_{us}X_{\tau\nu} = |X_{us}X_{\tau\nu}|\exp(i \delta_t).$

The decay angular distribution of the charge conjugate decay
$\tau^+\rightarrow \bar\nu_\tau +K^++\pi^0$ can be obtained from
above by reversing the sign of the four momenta, i.e.
$p_{1\mu}\rightarrow - p^\prime_{1\mu}$, etc., but keeping the
polarization vector unchanged. The strong interaction is invariant
under charge conjugation so $f_0$, $f_1$, $\delta_0$ and $\delta_1$
remain unchanged. The complex phase $\delta_t$ for causing $T$
violation changes sign because of hermiticity of the Hamiltonian and
this sign change also insures TCP invariance 
\cite{Tsai95a,Tsai96a,Tsai96b}.

Observations:
\begin{enumerate}
\item
Since $\cos(-\delta_t) = \cos(\delta_t)$, the interference term between
the $s$ wave part of $A_W$, represented by terms associated with
$|f_0|$ in Eq. (\ref{eq:2.16}), and the $A_X$ is not CP violating. Only
the $s$-$p$ interference represented by the term with $|f_1|$ in Eq.
(\ref{eq:2.16}) can have CP violation. [See also Appendix A, and 
\cite{Tsai95a,Tsai96a,Tsai96b}.]

\item 
The coefficients of terms proportional to
$\cos(\delta_0-\delta_1+\delta_t)$ is $T$ even because these terms
cannot violate CP in the limit $\delta_0-\delta_1=0$ according to the
TCP theorem. Similarly the coefficient of term proportional to
$\sin(\delta_0-\delta_1+\delta_t)$ is $T$ odd for similar reasons.

\item 
The purpose of the experiment is to find out whether CP is violated by
comparing the decay $\tau^-\rightarrow \nu_\tau + K^-+\pi^0$ with that
of CP conjugate decay $\tau^+\rightarrow \nu_\tau+K^++\pi^0$.
$\delta_t\ne 0$ or $\pi$ means  CP violation. From the $T$ even part of
Eq. (\ref{eq:2.16}) and that for $\tau^+$ decay, the CP violation is
proportional to
\begin{equation}
\cos(\delta_0-\delta_1+\delta_t)-\cos(\delta_0-
\delta_1-\delta_t)
= 2\sin(\delta_1-\delta_0)\sin\delta_t \label{eq:2.17}
\end{equation}
and for the $T$ odd term we have
\begin{equation}
\sin (\delta_0-\delta_1+\delta_t)-\sin(\delta_0-\delta_1-\delta_t)
= 2\cos (\delta_1-\delta_0)  \sin \delta_t \ . \label{eq:2.18}
\end{equation}
These two equations tell us three important things: 

\begin{itemize}
\item
CP violation comes from the imaginary part of the coupling constants,
i.e. $\sin\delta_t$, as expected.

\item
If there is no final state interactions, i.e. $\delta_1=\delta_0=0$,
only the $T$ odd part can have CP violation in agreement with the TCP
theorem.

\item
Near the $K^*$ resonance, we have $\delta_1\sim \frac{1}{2}\, \pi$ and
$\delta_0\sim 0$, hence the $T$ odd part hardly contributes to CP
violation whereas the $T$ even part contributes maximally near the
resonance. 
\end{itemize}

\item
We notice that other than the phase functions, the square of the $s$
wave part and $sp$ interference part of Eqs. (\ref{eq:2.15}) and
(\ref{eq:2.16}) have identical expressions. Only the square of the $p$
wave in Eq. (\ref{eq:2.15}) is unique there. These functions are
something like the relativistic version of Legendre polynomials and
thus universal in all similar calculations.

\item 
For the Tau-Charm Factory, the production of $\tau$ is more than 99\%\
$s$ wave \cite{Tsai95a}, hence the $\tau$ polarization is either along
the initial electron beam direction or opposite to it depending upon
the beam polarization. The $\tau$ angular distribution is almost
isotropic and the decay length is negligible. This means the production
angle must be integrated. In the $B$ factories, about 25\%\ of the cross
section is $d$ wave so polarization of the $\tau$ is only about 75\%\
even if the incident beam is 100\%\ polarized. On the other hand, in
the $B$ factories the energy of the $\tau$ is high enough to have
observable trajectory, so the production angle for $\tau^\pm$ need not
be integrated.

\item
The spin independent term $[2Q\cdot p_1-D(s+M_1^2)]$ in Eq.
(\ref{eq:2.16}) averages to zero when integrated with respect to the
direction of $p_1$. Thus it was ignored in Ref. \cite{Tsai95a}.
However, Mirkes and K\"uhn have pointed out \cite{Mirkes} that if we do
not integrate completely but only partially the phase space of $p_1$ we
can obtain a test of CP violation without using the polarized beam. In
the following we summarize their discovery. \end{enumerate}

\subsection{Test of CP violation in the semileptonic 
decay of $\tau$ without using polarized $\tau$}

\begin{figure}[htb]
\leavevmode
\centering
{\epsfxsize=2.9in\epsfysize=1.5in\epsfbox{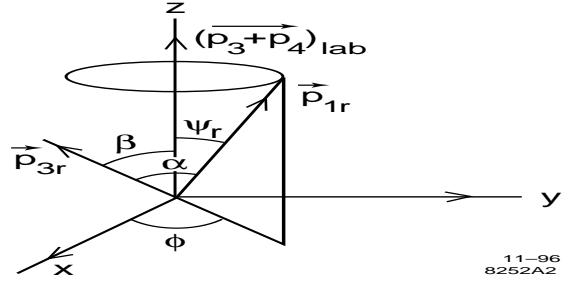}}
\caption{Definition of angles $\alpha$, $\beta$, and $\psi_r$ used in
Fig. 3.} 
\label{fig2} 
\end{figure}

\begin{figure}[htb]
\leavevmode
\centering
{\epsfxsize=2.9in\epsfbox{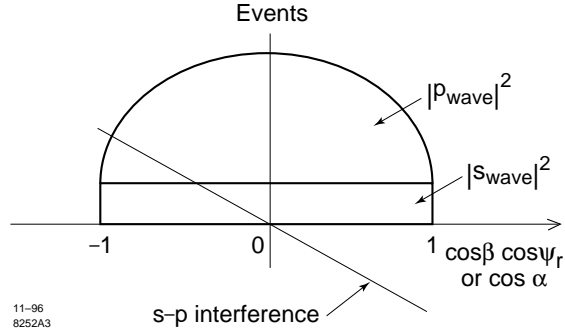}}
\caption[*]{Schematic illustration of the test of CP violation in tau
decay without using polarized $\tau$. CP violation is manifested by the
difference in slopes of the straight line labeled $s$-$p$ interference
between $\tau^-$ and $\tau^+$ decays.}
\label{fig3}
\end{figure}

In the rest frame of $p_3+p_4$ the term we are interested in (Eq.
(\ref{eq:2.16})) can be written as [see Fig. 2]
\begin{equation}
\frac{D(s+M^2_1)}{2} - Q\cdot p_1 = \vec Q_r\cdot \vec p_{1r} =
|\vec Q_r||\vec p_{1r}|
 \cos\alpha
\label{eq:2.19}
\end{equation}
where $\vec Q_r$ and $\vec p_{1r}$ are the 3 vector of $\vec Q$ and
$\vec p_1$ in the rest frame of $p_3+p_4$. We notice that
$\delta_0-\delta_1+\delta_t$ is a function of $s$ alone. So we hold $s$
fixed and plot the decay distribution as a function of $\cos\alpha$ if
the direction of $\tau$ is known. This distribution is different for
$\tau^+$ and $\tau^-$ decay if CP is violated. Thus one can obtain
$\delta_t$ by comparing the $\cos\alpha$ distributions of $\tau^+$ and
$\tau^-$ decay (see Fig. 3).

In the Tau-Charm Factory, the direction of $\tau^\pm$ is not
measurable. Mirkes and K\"uhn pointed out that even in this case one
can still measure $\delta_t$ by plotting the decay distribution as a
function of $\cos\beta\, \cos\psi_r$ where $\beta$ and $\psi_r$ are the
azimuthal angles of $\vec p_{3r}$ and $\vec p_{1r}$, respectively, with
the $z$ axis that is in the direction of $(\vec p_3+\vec p_4)_{\rm
lab}$ as shown in Fig. 2. Even though $\vec{p}_1$ is not measured
experimentally, its component along the $z$ direction in the laboratory
is confined along the surface of a cone around the $z$ axis, because
\begin{equation}
p^2_2 = 0 = (p_1-p_3-p_4)^2 = M^2_1+s-2(E_3+E_4)E_1+2p_{1z}
|\vec p_3+\vec p_4| \ . 
\label{eq:2.20}
\end{equation}
The $z$ component of $p_1$ in the rest frame of $p_3+p_4$ can be
obtained by boosting along the direction of $\vec z$ as shown in Fig.
2.
\begin{equation}
 p_{1rz} = |\vec p_{1r}| \cos\psi_r = - E_1\gamma\beta + \gamma p_{1z}
\label{eq:2.21}
\end{equation}
with
\begin{equation}
\gamma = \frac{E_3+E_4}{\sqrt s} \ , \qquad \beta\gamma =
\frac{|\vec p_3+\vec p_4 |}{\sqrt s} \ .
\label{eq:2.21a}
\end{equation}
Now $\cos\alpha$ in Eq. (\ref{eq:2.19}) is (see Fig. 2)
\begin{equation}
\cos \alpha = \cos\beta \cos\psi_r + \sin\beta \sin\psi_r\cos\phi \ .
\label{eq:2.22}
\end{equation}
After integrating with respect to $\phi$, the second term drops out and
thus by comparing the distributions of decay events as a function of
$\cos\beta\, \cos \psi_r$ for $\tau^+$ and $\tau_-$ we can find out if
CP is violated (see Fig. 3).

Physically, the CP violation in this case can be understood in the
following way:
\[ CP\, \vec Q_r (CP)^{-1} = - \vec Q^\prime,
\] and
\[
CP\, \vec p_{1r} (CP)^{-1} = - \vec p^\prime_{1r} ,
\]
where prime means the corresponding quantity for the charge conjugate
particle. Thus $CP\, \vec Q_r\cdot \vec p_{1r} (CP)^{-1} = \vec
Q^\prime_r\cdot \vec p^\prime_{1r}$.

This means that the coefficient of $\vec Q_r\cdot \vec p_1$ must be
equal to that of $\vec Q^\prime_r\cdot \vec p^\prime_{1r}$ if  CP
is conserved. When $p_1$ is not observed we simply replace it with the
$\phi$ averaged direction of $p_1$ as the $z$ axis, and instead of
plotting the events as a function of $\cos\alpha$, we plot the events
as a function of $\cos\beta\, \cos\psi_r$ in order to find out whether
CP is violated (see Fig. 3).

Since $\vec Q_r\cdot \vec  p_{1r}$ is $T$ even, we must have final
state interaction phases to have CP violation for this kind of test.

As this stage we might ask whether polarized $\tau^\pm$'s are still
needed for testing CP violation in $\tau$ decay. Let me give my opinion
on this subject:
\begin{enumerate}
\item
For a test of $T$ violation in $\tau^\pm \rightarrow
\nu^\tau+\mu^\pm+\nu_\mu$ the polarized $\tau$ is absolutely necessary
because $(\vec W_\tau\times\vec p_\mu)\cdot\vec W_\mu$ is the only $T$
violating quantity one can construct. This reaction involves only
leptons, thus if one finds CP violation in this reactions we can
conclude that a pure leptonic system violates CP.
\item 
If $\tau$ is polarized we can check CP violation using $W_3$, $W_4$ and
$\vec W\cdot(\vec p_3\times\vec p_4)$ terms. this is equivalent to
quadrupling the number of events compared with using angular asymmetry
discussed above.
\item
Since the polarization vector can be reversed, we can check whether the
CP violating effect is real or not by switching the sign of the
polarization.
\item
The overall production rate can be increased by a factor $(1+w_1w_2)$
where $w_1$ and $w_2$ are longitudinal polarization of $e^-$ and $e^+$
respectively (see Ref. \cite{Tsai95a}). 
\end{enumerate}

\section{SEMILEPTONIC DECAY OF HADRONS}

\begin{figure}[htb]
\leavevmode
\centering
{\epsfxsize=2.9in\epsfbox{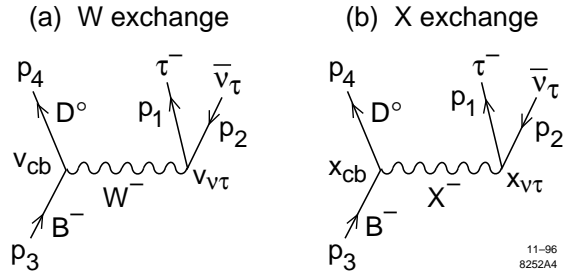}}
\caption[*]{Feynman diagrams for the reaction $B^-\rightarrow
D^0+\tau^-+\bar\nu_\tau$. $A_W$ and $A_X$ are explained in the caption
of Fig. 1.}
\label{fig4}
\end{figure}

This subject has been treated fully by many people in conjunction with
Weinberg's three Higgs doublet model 
\cite{Belanger,Garisto-Kane,Garisto95}. 
For tau decay we have final state interactions
but here we selected final states so that they have neither strong nor
electromagnetic final state interactions. Since there is no final state
interaction we have $\delta_1=\delta_0=0$ and thus
$\cos(\delta_1-\delta_0+\delta_t) = \cos(\delta_t) = \cos(-\delta_t)$
and hence we cannot have CP or $T$ violation coming from $T$ even terms
such as $S\cdot p_3$ and $S \cdot p_4$ or polarization independent
angular distributions. The only possible $T$ or CP violating term is
$\sin \delta_t Eps (p_1,S, p_3,p_4)$. In the rest frame of the lepton,
$p_1=M_1$, we have 
\[
Eps(p_1, S, p_3, p_4)=M_1 (\vec p_3 \times
\vec p_4) \cdot \vec S,
\]
where $\vec S$ is the direction of the spin of $p_1$.  When the lepton
is an unstable particle such as $\tau$ or $\mu$, the polarization can
be analyzed by measuring the energy angle distribution of a decay
particle and hence $S$ is replaced by the momentum of any of the decay
particles of $\tau$.

Let us consider the decay $B^\pm \rightarrow D^0+\tau^\pm+\nu_\tau$ as
an example. The Feynman diagrams for $W$ exchange and $X$ exchange are
shown in Fig. 4. We assume that the $X$ exchange contribution to be
much smaller than the $W$ exchange one, and thus we ignore $|A_X|^2$
compared with $|A_W|^2$ and the interference between the two. The
treatment here is similar to the previous section dealing with the tau
decay, except here we do not have to separate out the $p$ wave and $s$
wave parts in $A_W$ because we do not have to deal with the phase shift
due to the final state interactions.

\begin{equation}
\label{eq:C1}
A_W = V_{cb} J_\mu \bar u(p_1)\gamma_\mu(1-\gamma_5)v(p_2) 
\end{equation}
where
\begin{equation}
J_\mu = \VEV{D^0\,|\,\bar\psi_c(0) \gamma_\mu\psi_b(0)\,|\,B^-}
= (p_3-p_4)_\mu f_-(t)+(p_3+p_4)_\mu f_+(t) \label{eq:C2}
\end{equation}
where $p_1$, $p_2$, $p_3$, and $p_4$ are defined in Fig. 4 
and $t=(p_3-p_4)^2$.
\begin{equation}
A_X = X_{cb}X_{\tau\nu}
\VEV{D^0\,|\,\bar\psi_c(0)\psi_b(0)\,|,B^-} 
\bar u (p_1)(1-\gamma_5)v(p_2)
\ . \label{eq:C3}
\end{equation}
Now
\begin{eqnarray}
\VEV{D^0\,|\,\bar\psi_c(0)\psi_b(0)\,|,B^-} 
&=& \frac{(p_3-p_4)_\mu J_\mu}{m_c-m_b}= \frac{t}{m_c-m_b}\ f_-+
\frac{M^2_3-M^2_4}{m_c-m_b}\ f_+ \nonumber \\ &\equiv& f_0\
\frac{t}{m_c-m_b} \ .
\label{eq:C4}
\end{eqnarray}
The decay rate and decay distribution can be calculated by: 
\begin{eqnarray}
&&\Gamma(B^-\rightarrow D^0+\tau^-+\bar\nu_\tau) = \\ &=&
\frac{1}{2M_B}\,
\frac{1}{(2\pi)^5}\int \frac{d^3p_4}{2E_4} \int \frac{d^2p_1}{2E_1}\int
\frac{d^3p_2}{2E_2}\  \delta^4(p_3-p_4-p_1-p_2)
\left|\frac{g^2}{m^2_W}\, A_W+\frac{1}{m^2_X}\, A_X\right|^2 \ .
\nonumber
\label{eq:C5}
\end{eqnarray}
Let
\begin{eqnarray*}
V &=& f_- (\pslash_3-\pslash_4)+f_+(\pslash_3+\pslash_4) \ , \\[2ex]
V^* &=& f^*_- (\pslash_3-\pslash_4)
+ f^*_+ (\pslash_3+\pslash_4) \ . \end{eqnarray*}
We have
\begin{eqnarray}
&&\frac{A^+_WA_W}{2|V_{cb}|^2} =
\frac{{\rm Tr}}{4} \Big[(1+\gamma_5\Sslash)
(\pslash_1+M_1)V(1-\gamma_5) \pslash_2(1+\gamma_5)V^*\Big]
\nonumber \\  &=& |f_-|^2 \Big\{(S\cdot p_3)2M^3_1-(S\cdot
p_4)2M^3_1 +M^2_1(t-M^2_1)\Big\} 
\nonumber \\ && +|f_+|^2\Big\{(S\cdot p_3)2M_1\left[2p\cdot p_1
-t+4M^2_4\right]  
+ (S\cdot p_4)2M_1\left[2p\cdot p_1+t-4M^2_3\right] 
\nonumber \\ &&-4(p\cdot p_1)^2+4(p\cdot p_1)(M^2_3-M^2_4)
+ (M^2_1-t)(2M^2_3 + 2M^2_4-t)\Big\} 
\nonumber \\ && + {\rm Re}(f_-f^*_+)\Big\{ (S\cdot p_3)2M_1 
\left[2p\cdot p_1-t+M^2_1\right] 
+ (S\cdot p_4)2M_1\left[-2p\cdot p_1-t +M^2_1\right] 
\nonumber \\ && + 4M^2_1(-p\cdot p_1+M^2_3-M^2_4)\Big\} 
-8 {\rm Im}(f_-f^*_+)M_1 Eps(S,p_3,p_4,p_1) 
\label{eq:C6}
\end{eqnarray}
where $S$ is the direction of spin of $\tau$ and $p=p_3+p_4$.
\begin{eqnarray}
&& \frac{A^+_WA_X+A^+_XA_W}{\VEV{2\,|\,V_{cb}X_{cb}X_{\tau\nu}\,|\,
\frac{t}{m_c-m_b}}} =
\nonumber \\
&=& \frac{{\rm Tr}}{4}\, (1+\gamma_5\Sslash)(\pslash_1+M_1)
\Big[(1-\gamma_5)\pslash_2(1+\gamma_5)
V^*f_0+V(1-\gamma_5) \pslash_2(1+\gamma_5) f^*_0\Big] 
\nonumber \\ &=& \cos(\delta_t)  (f_0f^*_-)\Big\{(S\cdot p_3)4M^2_1
-(S\cdot p_4)4M^2_1+2M_1(t-M^2_1)\Big\}
\nonumber \\ &&+ \cos(\delta_t) (f_0f^*_+)
\Big\{2(S\cdot p_3)\left[2p\cdot p_1-t+M^2_1\right]
+2(S\cdot p_4) \left[-2p\cdot p_1-t+ M^2_1\right] 
\nonumber \\ && + 4M_1(-p\cdot p_1+M^2_3-M^2_4)\Big\}
- 8\sin(\delta_t)(f_0f^*_+) Eps(S, p_3, p_4, p_1) \ ,
\label{eq:C7}
\end{eqnarray}
where $p=p_3+p_4$ , $\delta_t $ is the CP violating phase in $A_X$ ,
i.e. $X_{cb}X_{\tau\nu} = |X_{cb}X_{\tau\nu}|\exp(i \delta_t)$. Since
there is no final state interaction we assume all form factors to be
real. Since only the relative phase between $A_W$ and $A_X$ is observable
we assume the KM matrix element to be real.

\section*{APPENDIX A---Comments on Nelson et al.'s \hfill\break
paper.}

In this appendix I would like to resolve two conflicting results in the
literature. Nelson et al. \cite{Nelson} discussed in a series of papers
the possibility of CP violation in the decay $\tau^\pm \rightarrow
\nu_\tau + \rho^\pm$ and my papers \cite{Tsai95a,Tsai96a,Tsai96b} which
showed that in the decay $\tau^\pm \rightarrow \nu_\tau+\pi^\pm+\pi^0$
CP violation can occur only through the interference of two Feynman
diagrams; one with $p$ wave final states, and the other with $s$ wave
final states for $\pi^\pm+\pi^0$. Since $\rho^\pm$ can have only $p$
wave final states, Nelson et al. must have made a mistake somewhere. I
first show that Nelson et al.'s result violates TCP and then show that
the reason for this violation is that their assumption of arbitrary 
amplitude for different helicity amplitudes for the decay $\tau^\pm
\rightarrow \nu_\tau+\rho^\pm$ violates rotational symmetry.

Let $A(h_{\rho^-}, h_\nu)$ be the helicity amplitude of $\tau^-
\rightarrow \nu_\tau+\rho^-$ in the rest frame of $\tau^-$ and
$B(h_{\rho^+}, h_{\bar \nu})$ be the corresponding amplitude for
$\tau^+\rightarrow \bar \nu_\tau+\rho^+$. If we ignore the final state
interaction between $\nu$ and $\rho$ we have 
\begin{equation} 
(TCP)\, A(h_{\rho^-}, h_\nu)(TCP)^{-1} = B(-h_{\rho^+},-h_{\nu^-}) \ .
\nonumber
\label{eq:A1} 
\end{equation} 
Nelson et al. proposed to measure
\begin{equation} 
\frac{A(-1,-\frac{1}{2})}{A(0,-\frac{1}{2})} = r_a =
|r_a| e^{i\beta_a} \label{eq:A2} 
\end{equation} 
and 
\begin{equation}
\frac{B(1,\frac{1}{2})}{B(0,\frac{1}{2})} = r_b = |r_b| e^{i\beta_b} \
. \label{eq:A3} 
\end{equation} 
They suggested that $|r_a|\ne |r_b|$ and $\beta_a \ne \beta_b$ would
indicate CP violation in $\tau^\pm \rightarrow \nu + \rho^\pm$.
Equation (\ref{eq:A1}) says that these two inequalities will also
violate TCP. Hence they are not viable tests of CP. TCP is violated
usually when any of the following three is violated: (1) Lorentz
invariance, (2) hermiticity of Hamiltonian, and (3) spin and statistics.

In the following we show that $r_a\ne r_b$ defined in Eqs.
(\ref{eq:A1}) and (\ref{eq:A3}) violates rotational symmetry that is
part of the Lorentz symmetry required for CPT and thus CPT is violated
in Nelson et al.'s papers.

For the left-handed and massless neutrino, the matrix element of
$\tau^-\rightarrow \nu_\tau+\rho^-$ can be written {\it uniquely} and
{\it covariantly} as
\begin{equation}
A(h_{\tau^-},h_\nu,h_{\rho^-})  = g\bar u(p_\nu,h_\nu)
\gamma_\mu
(1-\gamma_5)u(p_{\tau^-},h_\tau) \epsilon_\mu(h_{\rho^-})
\label{eq:A4}
\end{equation}
and similarly the matrix element of $\tau^+\rightarrow \bar\nu_\tau +
\rho^+$ can
be written as
\begin{equation}
A^\prime (h_{\tau^+},h_{\bar \nu}, h_{\rho^+})
= g^\prime \bar
u(p_{\tau^+},
h_{\tau^+})\gamma_\mu (1-\gamma_5) v (p_{\bar \nu},
h_{\bar\nu})\epsilon_\mu(h_{\rho^+}) \
. \label{eq:A5}
\end{equation}
If we ignore the final state interaction between $\rho$ and $\nu$, we
have from hermiticity $g^\prime = g^*$. If we assume further $T$
invariance then $g$ and $g^\prime$ must both be real and equal to each
other. The imaginary part of $g$ [or $g^\prime$] that causes $T$
violation cannot be measured unless there is another Feynman diagram
$B$ that is different in structure from $A$ interfering with $A$. For
if the structure is similar, we have $B=CA$ and thus 
\begin{equation}
A^+B+B^+A = (C+C^*)A^+A 
\label{eq:A6} 
\end{equation} 
and the phase of $g$ (or $g^\prime$) is unmeasurable.

Since Eq. (\ref{eq:A4}) is unique for $\tau^-$ decaying into
$\rho^-+\nu$, the diagram $B$ cannot have a different structure from
$A$, thus one cannot have $T$ violation in $\tau^-\rightarrow
\nu_\tau+\rho^-$. The diagram $B$ must contain $s$ wave for
$\pi^-+\pi^0$ system in order to avoid the cancellation of phases shown
in Eq. (\ref{eq:A6}).

In the rest frame of $\rho$, $\epsilon_\mu$ can have only three
components $\epsilon_x$, $\epsilon_y$, and $\epsilon_z$ with
$\epsilon_0=0$. We have thus
\begin{equation}
\epsilon_\mu\gamma_\mu = -\epsilon_z\gamma_z -
\left(\frac{\epsilon_x+i\epsilon_y}
{\sqrt 2}\right) \left(\frac{\gamma_x-i\gamma_y}{\sqrt 2}\right) 
-\left(\frac{\epsilon_x-i\epsilon_y}{\sqrt 2}\right)
\left(\frac{\gamma_x+i\gamma_y}{\sqrt
2}\right) \ . \label{eq:A7}
\end{equation}
$\epsilon_z\gamma_z$ contributes to $A(-\frac{1}{2},-\frac{1}{2},0)$
and the term $(\epsilon_x+i\epsilon_y)/\sqrt 2$
$(\gamma_x-i\gamma_z)/\sqrt 2$ contributes to
$A(+\frac{1}{2},-\frac{1}{2},1)$. The term
$(\gamma_x+i\gamma_y)/\sqrt2$ $(\epsilon_x-i\epsilon_y)/\sqrt2$ gives
zero because it projects out the right-handed neutrino. This shows that
the rotational invariance demands
\[ r_a=r_b\ .
\]
Thus $|r_a|\ne |r_b|$ and $\beta_a\ne \beta_b$ in Eqs. (\ref{eq:A2})
and (\ref{eq:A3}) violate rotational symmetry, hence this is not a
viable test of CP! The values of $r_a$ and $r_b$ depend upon the
Lorentz frame. Using explicit representation of spinors of $u$ and $v$
given in Ref. \cite{Tsai93}, one obtains in the rest frame of $\rho$:
\begin{equation} 
r_a = \frac{A\left(
\frac{1}{2},-\frac{1}{2},1\right)}
{A\left(-\frac{1}{2},-\frac{1}{2},0\right)} =
\frac{A^\prime\left(-\frac{1}{2},\frac{1}{2},-1\right)} 
{A^\prime\left( \frac{1}{2},\frac{1}{2},0 \right)} =  r_b
= -\sqrt 2\ \frac{E_\tau +M_\tau -p_\tau}{E_\tau+M_\tau +p_\tau} \ .
\label{eq:A8} 
\end{equation} 
We notice that the phase differences $\beta_a$ and $\beta_b$ are
meaningless quantities because the states with different helicities are
orthogonal to each other and thus they do not interfere, hence their
phase difference can never be measured! Usually the phase difference
depends upon the convention chosen and it is an irrelevant quantity.
For example, the $-$ sign in Eq. (\ref{eq:A8}) is due to my choice of
$(\epsilon_x-i\epsilon_y)/\sqrt 2$ to represent a $\rho$ with negative
helicity. Had I chosen $-(\epsilon_x-i\epsilon_y)/\sqrt2$, the negative
sign in Eq. (\ref{eq:A8}) would not be there. This shows that the phase
convention between different helicity states should not have any
observable consequence. Nelson et al.,'s statement that $\beta_a\ne 0$
or $\beta_b\ne 0$ implies a violation of $T$ invariance in the absence
of final state interaction  is thus erroneous.

I would also like to make a comment on the method of stage-two
spin-correlation (s2sc) advocated in Nelson et al.'s paper. The method
consists of calculating the density matrix of $\rho$ first and then
calculating the decay distributions of $\pi^-$ and $\pi^0$ from
$\rho^-$ decay using the helicity formalism. They claim using this
method one can measure $|r_a|$, $|r_b|$, $\beta_a$, and $\beta_b$ and
thus test CP violation. The method must be all wrong because $|r_a|\ne
|r_b|$ violates rotational symmetry and $\beta_a$ and $\beta_b$ are
meaningless quantities as shown above. Let me show why it is wrong.

The matrix element for the process
\begin{eqnarray*}
\tau^- \rightarrow \nu_\tau + &\rho^-& \\ 
&\srarrow& \pi^-+\pi^0
\end{eqnarray*}
can be written covariantly using the Breit-Wigner $\rho$ propagator:
\begin{equation}
g\, g_{\rho\pi}\bar u(p_2)\gamma_\mu(1-\gamma_5)u(p_1)\
\frac{g_{\mu\nu}-p_\mu p_\nu/p^2}
{(p^2-m^2_\rho)+i\Gamma m_\rho}\, q_\nu
\label{eq:A9}
\end{equation}
where $p=p_3+p_4$ and $q=p_3-p_4$ with $p_3=p_{\pi^-}$ and
$p_4=p_{\pi^0}$, $p_1=p_\tau$, and $p_2=p_\nu$. The rest of the
calculation follows the usual procedure of Feynman diagram calculation.
There is no need to introduce the density matrix for $\rho$, there is
no need to decompose Eq. (\ref{eq:A9}) into helicity amplitudes, and
the result is given essentially by the $p$ wave part of Eq.
(\ref{eq:2.15}), i.e. the coefficient of $|f_1|^2$ in Eq.
(\ref{eq:2.15}). Since the only place one can produce CP violation in
Eq. (\ref{eq:A9}) is the weak coupling constant $g$ being complex and
since it is squared in the cross section, there is no way CP violation
can be observed using Eq. (\ref{eq:A9}), we need interference with $X$
boson exchange to observe CP violation.

I would like to emphasize that the only way I know how to construct a
$T$ or CP violating theory without violating any sacred physics
principle is to have a complex coupling constant somewhere in the
theory. Hence all tests of $T$ and CP should have direct bearing on how
the proposed test can uncover this complex coupling constant. Had
Nelson et al. done this, they would have discovered that the tests they
have proposed have nothing to do with $T$ or CP violation.

Another remark I would like to make is pedagogical: one should avoid
doing the two-step calculation of calculating the density matrix of the
intermediate particle and then its decay as is done by Nelson et al. It
should be treated in one step with a propagator like Eq. (\ref{eq:A9}).
This way of treating the production of an unstable particle and its
subsequent decay was first done in Ref. \cite{Tsai65}.

We should also avoid unnecessary use of $D$ functions and Wigner
rotations occurring in the helicity formalism such as done in Nelson et
al.'s paper. Most of the problems in high energy physics can be done
covariantly. In Nelson et al.'s case, a problem that can be described
covariantly in two lines turned into a ten page Physical Review
nightmare with wrong results, because of the use of the two-step
process and a noncovariant helicity formalism.

In Appendix C we show that $\pi^\pm\pi^0$ must be in $I=2$ $s$ state
(Gell-Mann-Levy's $\sigma$ article) in the scalar exchange diagram. The
quark model of hadrons $\bar du$ and $u \bar d$ can only be in $I=1$
state and thus $I=2$ is forbidden by isospin conservation in the strong
interaction. Thus the decay mode, $\tau^\pm \rightarrow
\nu_\tau+\pi^\pm+\pi^0$ discussed in Ref. \cite{Tsai95a}, is not a very
good candidate for discovering the CP violation despite of its large
branching ratio.  To investigate whether the vertex $X_{ud}$ has an
imaginary part the best candidate is probably the decay process,
$\tau^\pm \rightarrow 3\pi + \nu$ that was discussed by Choi et al.
\cite{Choi}

I should mention that C. A. Nelson's paper was the first one to look
into CP violation in $\tau$  decay. Even though it proved to be wrong,
I am grateful for their effort because it interested me and caused me
to look into the problem.

\section*{APPENDIX B---Can $X^\pm$ boson have spin 1?}

The answer is no, because the axial and vector parts of the interaction
do not interfere in the final state. Similar to the $W^\pm$ boson, the
$W^{\prime \pm}$ boson can couple to quarks with either vector (V) or
axial vector (A) current if its spin is 1. Vector coupling can have
final states $J^P=0^+$ and $1^-$, whereas axial vector coupling can
have $J^P=0^-$ and $1^+$. For example, the final state $K^\pm+\pi^0$,
$J^P=0^-$ and $1^+$ are not possible, so only the vector current can
participate for both $W$ and $W^\prime$ and the rest of the final
strong interactions are identical in two cases. The matrix element for
$W^\prime$ must be proportional to that for $W$ for any particular
final state. Let $M_0$ be the CP conserving $W$ exchange diagram and
$M_1$ be the CP violating $W^\prime$ exchange diagram. Since for every
hadronic final state only either V or A contributes, we have
\begin{equation} M_1=cM_0 \ . 
\label{eq:B1} 
\end{equation} 
In this case
the interference between $M_0$ and $M_1$ is 
\begin{equation}
M^+_0M_1+M^+_1M_0 = (c+c^*)M^+_0M_0 
\label{eq:B2} 
\end{equation} 
and hence even if there is a CP violating complex phase in $M_1$, it
will not be observable. Thus no CP violation is observable in the
semileptonic decay of $\tau$ when $X$ is a spin 1 particle. For pure
leptonic decay, $W^\prime$ will not contribute to the CP violation as
long as neutrinos in the $W$ exchange have a definite helicity. In this
case only the portion of the $W^\prime$ exchange diagram which is
similar in structure to the $W$ exchange diagram will interfere with
the latter and again no complex CP violating phase in the $W^\prime$
exchange diagram could be detected for exactly the same reason as shown
in Eq. (\ref{eq:B2}). We conclude that we need to consider only spin 0
particle exchange interfering with the $W$ exchange diagram to produce
non-CKM-type CP violation.

In Ref. \cite{Tsai95b} it was shown that only spin 0 $X$ exchange
interfering with $W$ exchange can produce CP violation in pure leptonic
decay of $\tau$.

\section*{APPENDIX C---Can $2\pi$ decay mode of $\tau$ violate CP?}

The answer is yes, but very small because of quark model and isospin
conservation.

The final state $\pi^\pm+\pi^0$ from spin 0 $X$exchange must be in $s$
state and isospin $I=2$ because  of statistics. In the quark model the
scalar $X^\pm$ is coupled to $\bar ud$ or $\bar du$ which has $I=1$.
Since we are dealing with strong interactions after $\bar ud$ and $\bar
du$ are formed, the final state must sill have $I=1$. An $I=2$ state
cannot be obtained from an $I=1$ if isospin is conserved. Isospin is
violated by electromagnetic interactions and by mass difference between
$u$ and $d$ quarks. The mass difference between $\pi^\pm$ and $\pi^0$
is caused mostly by the latter. Since the mass difference between
$\pi^\pm$ and $\pi^0$ is about 3.5\%, we expect isospin conservation is
broken by a few percent. In principle, the imaginary part of of the
coupling constant $X_{ud}$ can also be obtained from CP violation in
the $\beta$ decay of $\pi^+$: $\pi^+\rightarrow \pi^0+e^+ +\nu$
\cite{MacFarlane}. We need to investigate further in detail how the
transversal polarization of a positron can be measured experimentally. 

\end{document}